\documentclass[aps,prl,
groupedaddress,amssymb,
noshowpacs,nofootinbib]{revtex4}

\usepackage{amsmath,amssymb}

\usepackage[usenames]{color}

\usepackage{graphicx}
\usepackage{mathptmx}
\usepackage{bbm}
\usepackage{bm}
\usepackage{times}

\newcommand{\beq}{\begin{eqnarray}}
\newcommand{\eeq}{\end{eqnarray}}

\makeatother

\begin{document}

\onecolumngrid

\title{A unitary renormalizable model of composite gravitons}

\author{\;Peter \surname{Orland}$^{ab}$}

\email{orland@nbi.dk, peter.orland@baruch.cuny.edu}

\affiliation{a. Baruch College, The 
City University of New York, 17 Lexington Avenue, 
New 
York, NY 10010, U.S.A. }

\affiliation{b. The Graduate School and University Center, The City University of New York, 365 Fifth Avenue,
New York, NY 10016, U.S.A.}

\begin{abstract}

A four-dimensional ${\rm SU}(4)$ confining Yang-Mills field, coupled to a fundamental fermion field and a bi-fundamental scalar field, has excitations with spin-2, but no other quantum numbers. These spin-2
excitations can be light or can condense, depending upon the scalar coupling. If condensation occurs, there is a massless spin-2 Goldstone boson 
with (possibly weakly) broken Lorentz invariance in the effective theory.  The low-lying spectrum contains additional spin-0 and spin-1 particles. We 
discuss how to couple these new fields to
other matter fields. To our knowledge, this is the only explicit proposal for a unitary and 
perturbatively-renormalizable local field theory of gravity. 

\end{abstract}


\maketitle


The nonrenormalizibility of Einstein's theory of gravity means that it must be modified at high energies. It is conceivable that in nature
the modification implies that the graviton is a bound state of more fundamental particles. We introduce a unitary and 
renormalizable model, whose spectrum contains 
two spin-2 excitations. The parameters of this model may be tuned, so that the bare mass-squared of at least one of the excitations 
is small or negative. The consequence is the appearance of either light massive gravitons, and/or massless 
gravitons as Goldstone bosons of spontaneously-broken Lorentz symmetry. The theorem 
of Weinberg and Witten forbids massless composite spin-2 excitations in a Lorentz-invariant theory with a conserved 
stress-energy-momentum tensor \cite{WW}. Thus, in the spontaneously-broken phase of our model, 
the spectrum cannot be exactly Lorentz invariant. 

 Historically, gravity theories with arbitrarily small graviton masses were thought ruled out
by experiment \cite{nogo}. Later it was shown this is not the case \cite{Vainshtein}; massive gravitons may 
be experimentally viable, provided the mass is sufficiently small.  A massive gravity theory should have 
an action similar to that of de Rham, Gabadadze and Tolley \cite{dRGT}, a construction without
Boulware-Deser ghosts \cite{BD}). The viability of a spontaneously-broken
phase depends upon the Lorentz symmetry breaking in the effective theory being sufficiently weak.

The study of composite gauge bosons began with the remarkable paper of Bjorken \cite{Bj}, who proposed that photons were Goldstone bosons of spontaneously-broken Lorentz invariance. Later this idea was extended to massless spin-2 Goldstone bosons \cite{4FermiGravitons}. The coupling of any massless spin-2 particles to matter is through the energy-momentum tensor \cite{Weinberg}. This leads naturally to Einstein's gravity as an effective theory. Banks and Zaks noticed that the vacuum of a model utilizing Bjorken's mechanism cannot be exactly Lorentz invariant \cite{BZ}, as Bjorken himself later observed \cite{Bj2}. A very general analysis
was done by Kraus and Tomboulis, showing that approximate Lorentz invariance can be present, including deSitter 
solutions \cite{KT}. They also pointed out that the effective Lagrangian for the gravitational Goldstone boson could not have a tadpole term, 
meaning that a large cosmological constant is absent. Kostelecky and Potting studied effective gravity theories based on spontaneous breaking of Lorentz symmetry 
more generally \cite{KP}. Unfortunately, the explicit models presented in References \cite{Bj,4FermiGravitons,Bj2, KT} have
nonrenormalizable four- or higher-fermion interactions. The spin-1 or spin-2 vacuum condensates are bilinears in the fermion field. One may 
hope that the long-range behavior of these models does not strongly depend on the cut-off at the critical coupling, but this seems unlikely (Kraus and Tomboulis suggested that models with 4-fermi interactions for photons or gravitons could be effective descriptions of  renomalizable gauge theories. To the author's knowledge, this has not been established, but see Reference \cite{KTREnorm?} for a large-$N$, strong-coupling analysis of a lattice gauge model).

In the light of the difficulty above, the approach using four-fermion interactions
does not improve upon
naively
quantized Einstein gravity, which Weinberg speculated to have an ``aymptotically-safe" critical point \cite{WAS}. The lack of perturbative 
renormalizability, however, hobbles the analytic
predictive power of such a theory, although numerical studies \cite{DynTriang} may eventually be successful. A serious concern is the appearance of higher-derivative, unitarity-violating, ultraviolet-divergent
counterterms
(the inducing of nonrenormalizable and higher-derivative terms is not pathological, provided the coefficients of these terms are finite).

Our unitary and renormalizable model is an ${\rm SU}(4)$ gauge theory, with fundamental fermions, henceforth called {\em aces}, with four types of charges, henceforth called {\em suits} 
(which may be labeled $\clubsuit, \;\diamondsuit,\; \heartsuit,\; \spadesuit$, and the four corresponding
antisuits ${\overline \clubsuit}, \;{\overline \diamondsuit},\; {\overline \heartsuit},\; {\overline \spadesuit}$). We denote suit
indices by lower case Roman letters, $a$, $b$, etc. The degrees of freedom are a traceless hermitian gauge field $A_{\mu}^{{\bar a},b}$, $\mu=0,1,2,3$, a fundamental four-component complex ace (Dirac) field $\psi_{a}$, ${\overline \psi}_{\bar a}$ and a 
bi-fundamental complex scalar field 
${\Phi}_{ab}$, ${\Phi}^{*}_{{\bar a}{\bar b}}$. The ace fields have no other quantum numbers. As we discuss below, the beta function of the gauge coupling is negative
(although those of the other couplings are not), and we expect confinement of aces into suit singlets, with a string tension proportional to the dimensional-transmutation scale.

In the confined phase, the lowest lying particle states are meson-like ace-antiace bound states and assemblies of four aces, analogous to baryons, which we call {\em quadyons}. We expect that the lightest quadyon is of orbital angular momentum zero. As a suit singlet it must be 
suit-antisymmetric and therefore spin-symmetric \cite{foot2}. Thus the lightest quadyon and the corresponding antiquadyon have spin-2. Note that the Weinberg-Witten theorem implies
that the spin-2 quadyon must be massive \cite{WW}, even if the constituent aces are massless. By additional (possibly charge-conjugation breaking) interactions with
bi-fundamental scalars, the quadyon and antiquadyon mix to produce a spin-2 condensate, with no quantum numbers except angular 
momentum. Four-fermion condensates in relativistic models were studied in References \cite{4Fconds}, although the context considered here is different.

Gravitons are not quadyons. There are distinct spin-2 quadyons and antiquadyons. Furthermore, if the dimensional-transmutation scale 
$\Lambda$ of the ${\rm SU}(4)$ gauge theory is large, the quadyons are too heavy to be gravitons; but if $\Lambda$ is small, there are many other light excitations which are not 
observed. We propose a mechanism by which quadyons and antiquadyons mix to 
produce light gravitons, with $\Lambda$ significantly larger than the scales of the Standard Model.

Other suit-singlet bound states can consist of one scalar with two aces, two scalars, and a scalar with an antiscalar. These, and meson-like 
bound states, ultimately couple to each other and gravitation, but not to other matter.

We briefly summarize our conventions. The diagonal Minkowski metric is $\eta_{\mu\nu}$, with $\eta_{00}=1, \eta_{11}=\eta_{22}=\eta_{33}=-1$. The elements of the basis of the Lie algebra $su$(4) will be denoted $t^{\alpha}, \alpha=1,\dots,15$, with normalization ${\rm Tr}\;t^{\alpha} t^{\beta}=\delta^{\alpha \beta}$ and structure coefficients $f^{\alpha \beta\rho}$, defined by $[t^{\alpha},t^{\beta}]={\rm i}f^{\alpha\beta\rho}t^{\rho}$. The gauge field
may be written as $A^{{\bar a}b}=\sum_{\alpha} A^{\alpha}t^{\alpha}_{{\bar a}b}$. We work with the standard Dirac matrices $\gamma^{\mu}$, satisfying 
$[\gamma^{\mu},\gamma^{\nu}]_{+}=2\eta^{\mu\nu}$, where $[,]_{+}$ is the anticommutator. The generators of proper orthochronous Lorentz transformations
are $\sigma^{\mu\nu}={\rm i}[\gamma^{\mu},\gamma^{\nu}]/4$.The Yang-Mills curvature tensor is
$F^{\alpha}_{\mu\nu} =\partial_{\mu}A^{\alpha}_{\nu}-\partial_{\nu}A^{\alpha}_{\mu}+
{\mathcal G}f^{\alpha\beta\gamma}A^{\beta}_{\mu}A^{\gamma}_{\nu}$, where, $\mathcal G$ is the gauge coupling and the summation convention for repeated indices is used. We will drop the bars over antisuit indices in the remainder of our discussion, for convenience.

We propose the Lagrangian for the gravitational sector:
\beq
{\mathcal L}\!\!&\!\!=\!\!&\!\!-\frac{1}{4} F^{\alpha}_{\mu \nu}F^{\alpha\;\mu\nu}+
{\overline \psi}_{a} ({\rm i} {\slash\!\!\!\!\!\!D}_{{a}b}-m\delta_{ab})\psi_{b} +\frac{1}{2} [(D_{\mu})_{ab;cd}{\Phi}_{cd}]^{*}
(D_{\mu})_{ab;jk}{\Phi}_{jk}+\frac{M^{2}}{2}{\Phi}^{*}_{ab}{\Phi}_{ab}
\nonumber \\
&+&\lambda^{abcdjklm}{\Phi}_{ab} {\Phi}_{cd}{\Phi}_{jk}{\Phi}_{lm}+{\lambda^{abcdjklm}}^{*}
{\Phi}^{*}_{ab} {\Phi}^{*}_{cd}{\Phi}^{*}_{jk}{\Phi}^{*}_{lm}
+Y{\Phi}_{ba} {\overline \psi}_{a} \psi^{C}_{b}-Y^{*}{\Phi}^{*}_{ab} {\overline \psi}^{C}_{a} \psi_{b}, \label{Lag1}
\eeq
where the covariant derivatives are 
\beq
{\slash\!\!\!\!\!\!D}_{ab}=\slash\!\!\!\!\partial \delta_{ab}
-{\rm i}{\mathcal G} \gamma^{\mu}A_{\mu\;a b}=\gamma^{\mu}\partial_{\mu} \delta_{ab}
-{\rm i}{\mathcal G} \gamma^{\mu}A_{\mu\;a b}, \label{dirac-cov}
\eeq
and 
\beq
(D_{\mu})_{ab;jk}{\Phi}_{jk}=(\partial_{\mu}\delta_{aj}\delta_{bk}-{\rm i}{\mathcal G} A_{\mu\;aj}\delta_{bk}-{\rm i}{\mathcal G} \delta_{aj}A_{\mu\;bk}){\Phi}_{jk}, \label{scalar-cov}
\eeq
the superscript $C$ denotes charge conjugation, defined on the ace field up to a phase $e^{{\rm i}\varphi_{C}}$:
\beq
\psi^{C}_{a}={\mathcal C} \psi_{a}{\mathcal C}^{-1}=e^{{\rm i}\varphi_{C}} C
{\overline \psi}_{a}^{T},\;\;
{\overline \psi}^{C}_{a}={\mathcal C}{\overline \psi}_{a}{\mathcal C}^{-1}=\gamma^{0}(\psi^{C}_{a})^{\dagger}, \label{charge-conj}
\eeq
where $C\gamma^{\mu}C^{-1}=-{\gamma^{\mu}}^{T}$ and complex constants $Y$ and $Y^{*}$ are Yukawa couplings and $\lambda^{abcdjklm}$ and ${\lambda^{abcdjklm}}^{*}$ project tensors of eight suits and eight antisuits, respectively, into singlets:
\beq
\lambda^{abcdjklm}=\sum_{P\in S_{8}} \lambda_{P}  \epsilon^{P(a)P(b)P(c)P(d)}\epsilon^{P(j)P(j)P(l)P(m)},\;\;
{\lambda^{abcdjklm}}^{*}=\sum_{P\in S_{8}} \lambda_{P}^{*}  \epsilon^{P(a)P(b)P(c)P(d)}\epsilon^{P(j)P(j)P(l)P(m)}, \label{scalar-couplings}
\eeq
where $P$ denotes a permutation in the symmetric group of eight objects $S_{8}$, and $\lambda_{P}$ is a coupling constant depending
on the permutation $P$. The number of independent complex values of $\lambda_{P}$ is $8!/(4!)^{2}$.
The Yukawa terms are gauge and Lorentz invariant, but violate charge conjugation and parity symmetries, unless $Y$ has vanishing 
real part, $Y=-Y^{*}$. 

In our discussion above of the Lagrangian (\ref{Lag1}), we have assumed a flat background spacetime with $g_{\mu\nu}=\eta_{\mu\nu}$, but finding generalizations with other background metrics is straightforward. In the this paper we consider only the flat-spacetime case in detail.

Under a gauge transformation $U\in {\rm SU}(4)$, $A_{\mu}\rightarrow U^{-1}A_{\mu}U+\frac{\rm i}{{\mathcal G}}U^{-1}\partial_{\mu}U$, $\psi \rightarrow
U\psi$, ${\overline \psi}\rightarrow {\overline \psi}U^{-1}$, ${\Phi}_{ab}\rightarrow U_{ac}U_{bd} {\Phi}_{cd}$, and 
${\Phi}^{*}_{ab}\rightarrow  {\Phi}^{*}_{cd}U^{-1}_{ca}U^{-1}_{db}$. The reader should be able to see that the quartic couplings in scalars, being suit-singlet operators, are 
${\rm SU}(4)$ gauge invariant.

Even if the ace mass $m$ vanishes, chiral tranformations $\psi_{a}\rightarrow e^{{\rm i}\alpha\gamma^{5}}\psi_{a}$, 
${\overline\psi}_{a}\rightarrow {\overline\psi}_{a}e^{{\rm i}\alpha\gamma^{5}}$, are not a symmetry of the Lagrangian (\ref{Lag1}). Not
only is this $U$(1) symmetry
broken by Yang-Mills configurations of nontrivial topological index \cite{U(1)}, but also by the Yukawa terms.


The gauge-field part of the Lagrangian (\ref{Lag1})
is asymptotically free. The one-loop contribution to the beta function is 
\beq
\beta({\mathcal G})=\frac{\partial {\mathcal G}}{\partial \ln\Lambda}=-\frac{11}{48\pi^{2}}N_{s}{\mathcal G}^{3}+\frac{1}{48\pi^{2}}N_{s}{\mathcal G}^{3}+ \frac{2}{48\pi^{2}}{\mathcal G}^{3}+\cdots
=-\frac{19}{24\pi^2} {\mathcal G}^{3}+\cdots,
\eeq
where the successive terms are contributions from gauge bosons, bi-fundamental scalars and aces, respectively, and the number of suits is $N_{s}=4$. As we stated above, we 
assume that suits are confined with physical masses proportional to $\Lambda$. The ace content of the quadyon field, which is the complex symmetric tensor $Q^{\mu\nu}(x)$, with engineering dimension one, is illustrated by the matrix element
\beq
\langle 0 \!\!\! \!& \!\!\!\! \vert  \!\!\! & \!\!\!Q^{\mu\nu}(x)\, \vert p_{1},s_{1},a_{1};p_{2},s_{2},a_{2},;p_{3},s_{3},a_{3};p_{4},s_{4},p_{4}\rangle_{\rm in}
=\Lambda^{-3}e^{{\rm i}x\cdot\sum_{j}p_{j}} \epsilon_{a_{1}a_{2}a_{3}a_{4}}\nonumber \\
&\!\!\times\!\!& [ (\gamma^{\mu}\kappa)_{s_{1}s_{2}} (\gamma^{\nu}\kappa)_{s_{3}s_{4}}
+ (\gamma^{\nu}\kappa)_{s_{1}s_{2}} (\gamma^{\mu}\kappa)_{s_{3}s_{4}}] \;K\left( \frac{p_{j}\cdot p_{k}}{\Lambda^{2}} \right)
, \label{quadwf}
\eeq
where $p_{1},\dots,\;p_{4}$, $s_{1},\dots,s_{4}$  and
$a_{1},\dots,a_{4}$ are the momenta, spinor and suit
indices, respectively, of the four aces, $\kappa$ is a $4\times 4$ matrix such that $\gamma^{\mu}\kappa$ and $\sigma^{\mu\nu}\kappa$ are symmetric \cite{McKeon}, and $K$ is a dimensionless Lorentz-invariant function of momenta. We emphasize that (\ref{quadwf}) only makes 
sense at momenta greater than $\Lambda$; otherwise the in-state ket must include the effects
of confinement, {\em e.g.}, virtual partons, nonabelian Wilson lines, etc. For simplicity, we have assumed that the ace masses are small, which is why we 
have included the factor of
$\Lambda^{-3}$ in (\ref{quadwf}), although our conclusions do not depend on this assumption.

The effective action for quadyons is induced by integrating out aces. This can be seen in 
the three-loop process shown in Figure 1.a and the six-loop process shown in Figure 1.b In the latter, quadyon-antiquadyon 
mixing occurs. The nonlocal quadyon-ace coupling is given by (\ref{quadwf}). We do not seriously suggest that these diagrams 
yield good approximations for these processes, but they illustrate how the effective quadyon action  
emerges. A numerical calculation may be possible by lattice methods. If the mass of the scalar particles, $M$ is much larger than
$\Lambda$, we can ignore mixing between quadyons and bound states containing scalars.

Processes like those in Figure 1.a and Figure 1.b lead to the terms of the effective Lagrangian for $Q_{\mu\nu}$ (suitably normalized), 
\beq
{\mathcal L}_{{\rm eff}\;1.a} = - {\mathcal A}Q_{\mu\nu}^{*}\, \Box^{\mu\nu}_{\;\;\alpha\beta}\, Q^{\alpha\beta}
+{\mathcal B}(Q_{\mu\nu}^{*} Q^{\mu\nu}-{Q^{\alpha}_{\;\;\;\alpha}}^{*} Q^{\beta}_{\;\;\;\beta}), \label{L1}
\eeq
and
\beq
{\mathcal L}_{{\rm eff}\;1.b} 
&=& 
-{\mathcal C} e^{-{\rm i}\delta} {Q_{\mu\nu}}^{*}\, \Box^{\mu\nu}_{\;\;\alpha\beta}\, {Q^{\alpha\beta}}^{*}
-{\mathcal C} e^{{\rm i}\delta} Q_{\mu\nu}\, \Box^{\mu\nu}_{\;\;\alpha\beta}\, Q^{\alpha\beta}
+{\mathcal D}     e^{-{\rm i}\frac{\chi}{2}}(Q_{\mu\nu}^{*} {Q^{\mu\nu}}^{*}-{Q^{\alpha}_{\;\;\;\alpha}}^{*} {Q^{\beta}_{\;\;\;\beta}}^{\!\!\!*}\, )
   \nonumber \\
&+&
{\mathcal D}     e^{{\rm i}\frac{\chi}{2}}(Q_{\mu\nu} Q^{\mu\nu}-Q^{\alpha}_{\;\;\;\alpha}Q^{\beta}_{\;\;\;\beta})
 , \label{L2}
\eeq
respectively, where ${\mathcal A}$, ${\mathcal B}$, ${\mathcal C}$, ${\mathcal D}$ are real and positive, $0\le\delta,\chi<2\pi$ and
\beq
 \Box^{\mu\nu}_{\;\;\alpha\beta}=(\delta_{\mu\alpha}\delta^{\nu\beta}-\eta_{\mu\nu}\eta^{\alpha\beta})\partial_{\lambda}\partial^{\lambda}
-2\delta^{\{ \stackrel{}{\mu}}_{\;\;\;\;\; \{\alpha}  \partial^{\stackrel{\nu}{}\}}\partial_{\stackrel{}{\beta}\}} +\eta_{\alpha\beta}\partial^{\mu}\partial^{\nu}
+\eta^{\mu\nu}\partial_{\alpha}\partial_{\beta}, \label{GravDiffOp}
\eeq
where the brackets $\{\}$ denote symmetrization of Lorentz indices. The mass terms, which in (\ref{L1}) are of order ${\mathcal B}$ and
which in (\ref{L2}) of are of order ${\mathcal D}$, have the only form consistent with the absence of
ghosts \cite{BD, Fierz, nogo}. Terms of second order in the quadyon field with more than two derivatives are induced, and
the additional derivatives
appear as $\partial_{\mu}/\Lambda$; we ignore these for now, but will return to them later. 

If $Q_{\mu\nu}$ is normalized so that the coefficients ${\mathcal A}$ and ${\mathcal C}$  are dimensionless, then
${\mathcal A}$ and ${\mathcal B}$ do not strongly depend on couplings in involving scalar fields, whereas ${\mathcal C}$ and ${\mathcal D}$ are proportional to 
$\lambda \vert Y\vert^{4}$ (we are not distinguishing the different scalar couplings $\lambda_{P}$, but
only considering rough dependence of the parameters in the effective Lagrangian). The effective Lagrangian is 
\beq
{\mathcal L}_{\rm eff}={\mathcal L}_{{\rm eff}\;1.a}+{\mathcal L}_{{\rm eff}\;1.b}+{\mathcal L}_{{\rm eff}\;2}(\{F\})+{\mathcal V}({Q^{\mu\nu}}^{*},Q^{\mu\nu}, \{F\}), \label{effLag}
\eeq
where ${\mathcal L}_{{\rm eff}\;2}(\{F\})$ is the Lagrangian for other bound states $\{F\}$, of the fundamental fields and the last term contains cubic- and higher-order parts of the potential as well as interactions with $\{F\}$. No ghosts are present in (\ref{effLag}), as our original system (\ref{Lag1}) is 
renormalizable and unitary.

Diagonalizing the quadratic form of the kinetic terms of  (\ref{effLag}), rescaling the fields appropriately, then diagonalizing the quadratic form
of the mass matrix yields 
\beq
{\mathcal L}_{\rm eff}&=&\sum_{\pm}\left[ -\frac{1}{2} H_{\pm\;\mu\nu}\, \Box^{\mu\nu}_{\;\;\alpha\beta}\, {H_{\pm}}^{\alpha\beta}
+ \frac{1}{2}m_{\pm}^{2}\, (H_{\pm\;\mu\nu} {H_{\pm}}^{\mu\nu}-{H_{\pm}}^{\alpha}_{\;\;\;\alpha} {H_{\pm}}^{\beta}_{\;\;\;\beta}) \right]
+{\mathcal L}_{{\rm eff}\;2}(\{F\})+{\mathcal V}({H_{+}}^{\mu\nu},H_{-}^{\mu\nu}, \{F\}) ,
\eeq
where ${H_{\pm}}^{\mu\nu}$ are real symmetric tensor fields with masses
\beq
m^{2}_{\pm}=\frac{{\mathcal A}{\mathcal B}+{\mathcal C}{\mathcal D}\cos(\delta-\chi)}{{\mathcal A}^{2}-{\mathcal C}^{2}}\pm
\sqrt{ \left[ \frac{{\mathcal A}{\mathcal B}+{\mathcal C}{\mathcal D}\cos(\delta-\chi)}{{\mathcal A}^{2}-{\mathcal C}^{2}} \right]^{2}-\frac{{\mathcal B}^{2}-{\mathcal D}^{2}}{{\mathcal A}^{2}}       }. \label{MEigen}
\eeq
The excitations of the fields $H_{\pm}^{\mu\nu}$ have spin equal to two, but no other quantum numbers.

In principle, the phases of the original model 
(\ref{Lag1}) are: i) $m^{2}_{+}>0, m^{2}_{-}>0$, ii) $m^{2}_{+}>0, m^{2}_{-}<0$, iii) $m^{2}_{+}<0, m^{2}_{-}>0$ and iv) $m^{2}_{+}<0, m^{2}_{-}<0$. The Weinberg-Witten theorem
forbids $m^{2}_{+}$ or $m^{2}_{-}$ being exactly equal to zero \cite{WW}. Thus any phase boundary is a hypersurface 
of first-order transitions, with no critical points. For sufficiently large $\lambda\vert Y\vert^{4}$, we will have ${\mathcal B}<{\mathcal D}$, and either phase ii) or iii) occurs. For small $\lambda \vert Y\vert^{4}$, phase i) must exist. Only a nonperturbative calculation of 
${\mathcal A}$, ${\mathcal B}$, ${\mathcal C}$, ${\mathcal D}$ and $\delta-\chi$, can determine whether Phase iv) exists in our model. It is
clear, however, that Phase i) and either Phase ii) or Phase iii) (or both) are possible. In the remainder of this letter, we consider only these possibilities.

Suppose that one of the two symmetric tensor fields condenses. Without loss of generality, we assume $\langle H_{+\;\mu\nu}\rangle=0$ and $\langle H_{-\;\mu\nu}\rangle\neq 0$, {\em i.e.}, Phase ii). The proper orthochronous subgroup $SO(1,3)$ of the Lorentz group is thereby spontaneously broken to the trivial group. A coordinate transformation is making $\langle H_{-\;\mu\nu}\rangle$ diagonal and constant over spacetime exists.

The Goldstone field $H_{-\;\mu\nu}$ has six components, only two of which are the helicity states of the graviton \cite{KT, KP}.
The potential ${\mathcal V}$ does not depend on these components. The effective Lagrangian in $H_{-\;\mu\nu}$ may have terms with more than two derivatives, but no non-derivative terms can be present.
By general arguments \cite{generalarguments}, the long-range form of the effective action of $H_{-\;\mu\nu}$ is the Einstein-Hilbert
action, coupled to dark suit-singlet effective fields ($H_{+\;\mu\nu}$, bound states with scalars or suit-antsuit pairs are among these).

We next discuss coupling to other fields. Coupling suitless two-component fermions $\Psi_{\mathfrak A}$ requires the existence of suited two-component fermions $\Pi_{\mathfrak Aab}$, coupling through new Yukawa terms in the Lagrangian,
\beq
 {\mathcal L}_{2}=y\,{\overline \Psi}_{{\mathfrak A}} \Phi_{ab}^{*}
\Pi_{\mathfrak Aab}+y^{*}\,{\overline \Pi}_{{\mathfrak A}ab} \Phi_{ab} \Psi_{\mathfrak A} .   \label{Lag2}
\eeq
If the fermi fields included in (\ref{Lag2}) are coupled to a gauge field $B_{\mu\;{\mathfrak A}{\mathfrak B}}$, there are further terms:
\beq
{\mathcal L}_{3}={\overline \Psi}_{\mathfrak A} {\rm i}\; {\slash\!\!\!\!\!{D}}_{{\mathfrak A}{\mathfrak B}}\Psi_{\mathfrak B}+
{\overline \Pi}_{{\mathfrak A}ab} ({\rm i}\; {\slash\!\!\!\!\!{D}}_{{\mathfrak A}ab;{\mathfrak B}cd}-\mu_{{\mathfrak A}ab;{\mathfrak B}cd} )\Pi_{{\mathfrak B}cd}
\label{Lag3}
\eeq
where $\mu_{{\mathfrak A}ab;{\mathfrak B}cd}$ is the mass
matrix of $\Pi_{{\mathfrak A}ab}$, and the differential operators in (\ref{Lag3}) are
\beq
{\slash\!\!\!\!\!{D}}_{{\mathfrak A}{\mathfrak B}}=\slash\!\!\!\!\partial \delta_{{\mathfrak A}{\mathfrak B}}
-{\rm i}g {\slash\!\!\!\!\!{B}}_{{\mathfrak A}{\mathfrak B}},
\;{\rm and}\;\;
 {\slash\!\!\!\!\!{D}}_{{\mathfrak A}ab;{\mathfrak B}cd}=\slash\!\!\!\!\partial \delta_{{\mathfrak A}{\mathfrak B}}\delta_{ac}\delta_{bd}
-{\rm i}g {\slash\!\!\!\!\!{B}}_{{\mathfrak A}{\mathfrak B}}\delta_{ac}\delta_{bd} -{\rm i}{\mathcal G} {\slash \!\!\!\!\! A}_{ac}\delta_{bd}
-{\rm i}{\mathcal G}\delta_{ac}\, {\slash \!\!\!\!\! A}_{bd},
\eeq
where $g$ is the gauge coupling associated with $B_{\mu\;{\mathfrak A}{\mathfrak B}}$.

A two-component Higgs field $\phi_{\mathfrak A}$ can also be coupled to $\Phi_{ab}$ through a 
quartic term,
\beq
{\mathcal L}_{4}=\Omega \phi_{\mathfrak A}^{*}\phi_{\mathfrak A} \Phi_{ab}^{*}\Phi_{ab}\;.
\eeq

Thus, elementary quarks, leptons and the Higgs can be coupled to our model of gravity through ${\mathcal L}={\mathcal L}_{1}+{\mathcal L}_{2}+{\mathcal L}_{3}+{\mathcal L}_{4}.$ Unfortunately, we have no direct renormalizable coupling to the Standard Model gauge fields, in particular the gluon field, which possesses most of the mass of visible matter. Furthermore, for the case of spontaneous Lorentz symmetry breaking, we cannot expect that such a coupling results from the arguments of Reference \cite{Weinberg}, because Lorentz invariance is not exact in the effective theory. To
couple gravitons to all the fields of the Standard model may require grand unification, such as embedding the gauge symmetry
${\rm SU}(4)\times {\rm SU}(3)\times {\rm SU}(2)\times {\rm U}(1)$ into a larger gauge group, such as SU($9$). Work on this idea is in progress.

At sufficiently high temperature, there is a deconfining phase transition. Our argument for spin-2 bound states of aces should remain valid in
the high-temperature phase, although it is not clear whether light gravitons are present in this phase. This phase transition should have implications for 
the early universe.

Our model is formulated with a background metric (taken here to be the Minkowski metric). It is conceivable that our model of gravity could be at least approximately background independent, if our quantum field theory is asymptotically topological in the infrared. This would mean that after integrating out fast modes of the functional integral with Lagrangian (\ref{Lag1}), \`a la Wilson, the result is a topological quantum field theory. It is unclear to the author whether this is the case.

\begin{center}

{\bf Acknowledgements}
\end{center}

\begin{acknowledgments}
During the long gestation of the ideas presented here, I discussed them with J. Ambj{\o}rn, C.F. Kristjansen, 
V.P. Nair, N. Obers, G.W. Semenoff, P. Woit and C.K. Zachos (none of whom bear responsibility for any misconceptions). This 
work is supported in part by a grant from the PSC-CUNY.

\end{acknowledgments}

\newpage

\thispagestyle{empty}

\onecolumngrid

\begin{center}

\begin{picture}(100,110)(-25,0)

\thicklines

\put(-5, 68){\large(a)}

\put(-15,94){${Q_{\mu\nu}}$}
\put(58,94){${Q_{\alpha\beta}}^{*}$}

\put(0,90){\oval(5,30)}
\put(50,90){\oval(5,30)}

\put(-13,90){\oval(3,3)[t]}
\put(-10,90){\oval(3,3)[b]}
\put(-7,90){\oval(3,3)[t]}
\put(-4,90){\oval(3,3)[b]}

\put(54,90){\oval(3,3)[t]}
\put(57,90){\oval(3,3)[b]}
\put(60,90){\oval(3,3)[t]}
\put(63,90){\oval(3,3)[b]}

\put(2.5,100){\line(1,0){45}}
\put(2.5,93){\line(1,0){45}}
\put(2.5,86){\line(1,0){45}}
\put(2.5,79){\line(1,0){45}}

\put(2.5,100){\vector(1,0){23}}
\put(2.5,93){\vector(1,0){23}}
\put(2.5,86){\vector(1,0){23}}
\put(2.5,79){\vector(1,0){23}}


\put(-5, 18){\large(b)}

\put(-15,44){${Q_{\mu\nu}}$}
\put(59,44){${Q_{\alpha\beta}}$}

\put(0,40){\oval(5,30)}
\put(50,40){\oval(5,30)}

\put(-13,40){\oval(3,3)[t]}
\put(-10,40){\oval(3,3)[b]}
\put(-7,40){\oval(3,3)[t]}
\put(-4,40){\oval(3,3)[b]}

\put(54,40){\oval(3,3)[t]}
\put(57,40){\oval(3,3)[b]}
\put(60,40){\oval(3,3)[t]}
\put(63,40){\oval(3,3)[b]}

\put(2.5,50){\line(1,0){45}}
\put(2.5,43){\line(1,0){45}}
\put(2.5,36){\line(1,0){45}}
\put(2.5,29){\line(1,0){45}}

\put(2.5,50){\vector(1,0){10}}
\put(2.5,43){\vector(1,0){10}}
\put(2.5,36){\vector(1,0){10}}
\put(2.5,29){\vector(1,0){10}}

\put(47.5,50){\vector(-1,0){10}}
\put(47.5,43){\vector(-1,0){10}}
\put(47.5,36){\vector(-1,0){10}}
\put(47.5,29){\vector(-1,0){10}}


\multiput(30,30)(0.5,1.5){22}{\circle*{0.6}}
\multiput(25,37)(0.75,1.25){22}{\circle*{0.6}}
\multiput(20,44)(1.1,1){20}{\circle*{0.6}}
\multiput(17,51)(1.9,1){12}{\circle*{0.6}}

\put(35,45){\vector(1,3){1}}
\put(35.5,46.5){\vector(1,3){1}}

\put(28.5,43){\vector(2,3){2}}
\put(29.75,45){\vector(2,3){2}}

\put(29.,52.){\vector(1,1){2}}
\put(30.5,53.5){\vector(1,1){2}}

\put(27.,56.25){\vector(2,1){1}}
\put(29,57.25){\vector(2,1){1}}


\put(29.5,29){\circle*{2}}
\put(24.5,36){\circle*{2}}
\put(19.5,43){\circle*{2}}
\put(16,50){\circle*{2}}
\put(41,63){\circle*{2}}

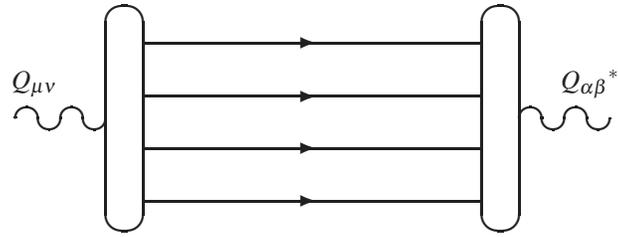
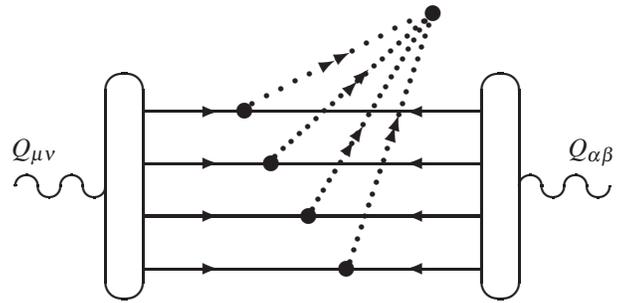
\put(-15,10){Figure 1: In (a) the quadyon is shown, moving from} 
\put(-15,6){left to right, as a bound state of four fermions. In (b), } 
\put(-15,2){each fermion emits a scalar (dotted line with double }
\put(-15,-2){arrows), becoming an antifermion, producing mixing} 
\put(-15,-6){of quadyons and antiquadyons.}

\end{picture}

\end{center}


\begin{thebibliography}{xx}
\bibitem{foot2} Note the analogy in chromodynamics with the $\Delta^{++}$, which is symmetric in flavor and antisymmetric in color, hence is spin symmetric.
\end{thebibliography}
\end{document}